\documentclass{czjphys}         
\usepackage{amsmath}
\usepackage{amssymb}
\usepackage{cite}

 \def\k{\kappa}

\def\rriga{\kern-4pt-\kern-4pt - \kern-4pt -}

\def\({\left(}\def\){\right)}

\def\half{{\textstyle{\frac{1}{2}}}}

\def\riga{-\kern-5pt - \kern-5pt -}

 \def\I{{\bar 1}}
\def\a{\alpha}

\def\d{\delta}
\def\g{\gamma}

\def\D{\Delta} \def\L{\Lambda}
\def\bac{{C\kern-5.5pt I}}
\def\bbc{{C\kern-8pt I}}

\def\be{\begin{equation}}
\def\eqn{\begin{equation}\label}

\def\bea{\begin{eqnarray}}
\def\eqnn{\bea\label}
\def\eea{\end{eqnarray}}
\def\nn{\nonumber}

\newcommand{\eqna}[1]{\begin{subequations} \label{#1}
\begin{eqnarray}}
\def\eena{\end{eqnarray}
\end{subequations}}

\def\vr{\vert} \def\bbr{I\!\!R}

\def\nt{\noindent}

\def\ca{{\mathfrak a}}
\def\cg{{\mathfrak g}} 
\def\ck{{\mathfrak k}} \def\cc{{\mathfrak C}}

\begin{document}

\title{On the Unitarity of D=9,10,11 Conformal Supersymmetry}

\authori{V.K. Dobrev}
\addressi
{School of Informatics,
University of Northumbria,\\ Newcastle-upon-Tyne, UK\\
and\\ Institute of Nuclear Research and Nuclear Energy,\\
Bulgarian Academy of Sciences, Sofia, Bulgaria (permanent
address)}

\authorii{A.M. Miteva}     \addressii{
Department of Physics, Sofia University, Sofia, Bulgaria}

\authoriii{R.B. Zhang}
\addressiii{School of Mathematics and Statistics,\\
University of Sydney,\\ Sydney, New South Wales, Australia}

  \authoriv{B.S. Zlatev}
  \addressiv{Institute of Nuclear Research and Nuclear Energy,\\
Bulgarian Academy of Sciences, Sofia, Bulgaria}

\authorv{}      \addressv{}
\authorvi{}     \addressvi{}

\headauthor{Dobrev et al.}            
\headtitle{On the Unitarity of D=9,10,11 Conformal Supersymmetry}
\lastevenhead{Dobrev et al.:  \ldots} 

\refnum{A}
\daterec{XXX}    
\issuenumber{0}  \year{2001}
\setcounter{page}{1}

\maketitle
\begin{abstract}
We consider the unitarity of $D=9,10,11$
conformal supersymmetry
using the recently established classification of the UIRs
 of the superalgebras ~$osp(1|2n,\bbr)$.
\end{abstract}


\section{Introduction}

Recently,  applications in string theory require the
knowledge of the UIRs of the conformal superalgebras for ~$D>6$.
In these applications most prominent role play the superalgebras
$osp(1\vr\,2n)$, cf., e.g.,
\cite{Tow,HW,PHNW,Gun,Luk,BvP,DFLV,Ban,SWK,BCB}.
Initially, the superalgebra $osp(1\vr\,32)$
was put forward for $D=10$ \cite{Tow}. Later it was realized that
$osp(1\vr\,2n)$ would fit any dimension, though they are minimal
only for $D=3,9,10,11$ (for $n=2,16,16,32$, resp.) \cite{DFLV}.
In all cases one needs to find first the UIRs of
~$osp(1\vr\, 2n,\bbr)$. This was done for general $n$ in \cite{DoZh}.
In the present paper we consider the implications for
conformal supersymmetry for $D=9,10,11$.

\section{UIRs of the superalgebras  $\ osp(1\vr\, 2n,\bbr)$}

\nt
The conformal superalgebras in $D=9,10,11$ are ~$\cg ~=~
osp(1\vr\, 2n,\bbr)$, $n=16,16,32$, resp., cf. \cite{Tow,DFLV}.
The even subalgebra of ~$osp(1\vr\, 2n,\bbr)$~ is the algebra
~$sp(2n,\bbr)$ with maximal compact subalgebra
~$\ck = u(n) \cong su(n) \oplus u(1)$.
We label the relevant representations of ~$\cg$~
by the signature:
\eqn{sgn}\chi ~=~ [\, d\,;\,a_1\,,...,a_{n-1}\,] \end{equation}
where ~$d$~ is the conformal weight, and ~$a_1,...,a_{n-1}$~
are non-negative integers which are Dynkin labels of the
finite-dimensional UIRs of the subalgebra $su(n)$ (the simple
part of ~$\ck$).
The positive energy UIRs of ~$\cg$~ for any $n>1$
are given in the following list  \cite{DoZh}:
\eqnn{unt}
&&d ~\geq~ d_{1} ~=~ n -1 + \half(a_1 + \cdots + a_{n-1})
\ , \quad {\rm no\ restrictions\ on}\ a_j \ , \\
&&d ~=~ d_{12} ~=~ n - 2 + \half(a_2 + \cdots + a_{n-1} +1)
\ , \quad a_1 ~=~ 0 \ , \nn\\
&& ... \nn\\
&&d ~=~ d_{j-1,j} ~=~ n -j + \half(a_j + \cdots + a_{n-1} +1)
\ , \quad a_1 ~=~ ... ~=~ a_{j-1} ~=~ 0 \ , \nn\\
&& ... \nn\\
&&d ~=~ d_{n-1,n} ~=~ \half
\ , \quad a_1 ~=~ ... ~=~ a_{n-1} ~=~ 0 \ .\nn \eea
These UIRs are realized as the irreducible quotients of Verma
modules $V^\L$ of lowest weight $\L=\L(\chi)$. The weight is
fixed by its products with the simple roots $\a_i$
($i=1,\ldots,n$) of $\cg$~\cite{DoZh}:
\eqna{rrr} 
&&(\L , \a_k^\vee ) ~=~
(\L , 2\a_k/(\a_k,\a_k) ) ~=~ (\L , \a_k) ~=~
-\, a_k \ , ~~~ k<n\ , \\
&&(\L , \a_n^\vee ) ~=~  2 (\L , \a_n) ~=~
2 d + a_1 + \cdots + a_{n-1}
\ . \eena

\section{Unitarity for the conformal subalgebras}

\nt {}From the prescription of \cite{DFLV} follows that the even
subalgebra ~$sp(2n,\bbr)$, $n=16,16,32$, resp., contains the
conformal algebra ~$\cc ~=~ so(D,2)$, $D=9,10,11$. Then ~$\ck$~
contains the maximal compact subalgebra ~$so(D)\oplus so(2)$~ of
~$\cc$, $so(2)$ being identified with the $u(1)$ factor of ~$\ck$,
and ~$su(n)$~ contains the algebra ~$so(D)$. The easiest way to
describe the embeddings is via the root systems. The superalgebra
$\cg$ is the split real form of the basic classical superalgebra
$osp(1\vr\, 2n)$ and has the same root system. The root system of
~$su(n)$, actually of ~$sl(n)$, is comprised of the even simple
roots of ~$\cg$~: ~$\a_i\,, i=1,\ldots,n-1$, with standard non-zero
products: ~$(\a_i\,,\a_i) =2$, ($i=1,\ldots,n-1$),
~$(\a_i\,,\a_{i+1}) =-1$, ($i=1,\ldots,n-2$). The root system of
~$so(D)$~ is comprised of simple roots ~$\g_j\,$,
$j=1,\ldots,\ell\equiv[D/2]$.

For even $D$ the non-zero scalar products are:
~$(\g_j\,,\g_j) ~=~ 2\k$, ($j=1,\ldots,\ell$),
~$(\g_j\,,\g_{j+1}) ~=~ -\k$, ($j=1,\ldots,\ell-2$),
~$(\g_{\ell-2}\,,\g_{\ell}) ~=~ -\k$,
where $\k$ is a non-zero common multiple (it is inessential since it
cancels in the Cartan matrix elements).
In the case of interest ~$D=10$, ~$\ell=5$, ~$n=16$~:
\eqnn{bas10}
&&\g_1 ~=~ {\a_4 + \a_7 + \a_9 + \a_{12}} \ ,\cr 
&& \g_2 ~=~ {\a_3 + \a_6 + \a_{10} + \a_{13}}\ , \cr
&& \g_3 ~=~ {\a_2 + \a_4 + \a_{5} + \a_{6}
+ \a_7 + 2\a_{8} + \a_{9}
+ \a_{10} + \a_{11} + \a_{12}+ \a_{14}}\ , \cr
&&\g_4 ~=~ {\a_3 + \a_5 + \a_{7} + \a_{15}} \ ,\cr 
&&\g_5 ~=~ {\a_1 + \a_9 + \a_{11} + \a_{13}} \ . \eea
(The roots $\g_i$ satisfy the prescribed products with $\k=4$.)
Correspondingly, the $so(10)$ Dynkin labels
~$r_k ~\equiv~ -\k (\L , \g_k^\vee) ~=~-(\L , \g_k)$~ are:
\eqnn{basm10}
&& r_1 ~=~ {a_4 + a_7 + a_9 + a_{12}} \ , \cr
&& r_2 ~=~ {a_3 + a_6 + a_{10} + a_{13}}\ , \cr
&& r_3 ~=~ {a_2 + a_4 + a_{5} + a_{6}
+ a_7 + 2a_{8} + a_{9}
+ a_{10} + a_{11} + a_{12}+ a_{14}}\ , \cr
&&r_4 ~=~ {a_3 + a_5 + a_{7} + a_{15}} \ , \cr
&&r_5 ~=~ {a_1 + a_9 + a_{11} + a_{13}} \ . \eea
The dimensions of the $so(10)$ UIRs is:
\eqn{dim10}
{\rm \dim}~L(r_1,...,r_5) ~ = ~ \prod_{1\leq s<t \leq 5}
\frac{ n^2_t - n^2_s}{ (t -s) \ (t+s-2) } \ ,
\end{equation}
where we use the additional parametrization:
\eqnn{param} &n_1 ~=~ \half (r_5 - r_{4}) \ , \qquad
n_2 ~=~ \half (r_5 + r_{4}) +1\cr
&n_s ~=~ \half (r_5 + r_{4}) + r_{3} + \dots +
r_{6 - s} + s-1\ , \qquad s=3,...,5 \ .\eea
The parameters $n_s$ are either all
integer or all half-integer obeying:
\eqn{} n_5 ~>~ n_{4} ~>~ n_3 ~>~ n_2 ~>~ \vr n_1\vr ~\geq~ 0\ .
\end{equation}

It is  known that the unitarity restrictions for a conformal
superalgebra $\ca$ are stronger than those for the even subalgebra
of $\ca$ (for $D=4,6$ cf. \cite{DP}). Here, in addition the
conformal algebra $so(D,2)$ is smaller than the even subalgebra
$sp(2n)$. Thus, the unitarity conditions for $so(D,2)$ are given
only in terms of ~$r_i\,$. Thus, $so(D,2)$ unitarity would not
require that all parameters $a_k$ are non-negative integers - that
would be required only for their combinations ~$r_i\,$. The
unrestricted parameters $a_k$ are combined in so-called tensorial
charges \cite{Tow}. We shall leave this together with more detailed
analysis for a follow-up paper.

Here, due to the lack of space, we only consider briefly the
reduction of the the fundamental irreps  of $su(n)$ to
$so(D)$  irreps.  We list only the main $so(D)$ component with
signature following directly from the embedding formulae, e.g.,
\eqref{basm10} for $so(10)$.

Of course, the one-dimensional irreps, when ~$a_i=0=r_s$~ for all
~$i,s$, coincide.  In the Dynkin labeling of the $n-1$ fundamental
irreps ~$\L_k$~ of $su(n)$ are characterized for fixed $k$ by ~$a_i
~=~ \d_{ik}\,$. The  fundamental irreps ~$\L^o_t$~ of $so(D)$
are characterized for fixed $k$ by ~$r_s ~=~ \d_{st}\,$. The
16-dimensional fundamental ~$su(16)$~ UIR with ~$a_1=1$
 gives the fundamental 16-dimensional $so(10)$ spinor
when ~$r_5=1$,  while the conjugated 16-dimensional fundamental
~$su(16)$~ UIR with ~$a_{15}=1$ gives the conjugated 16-dimensional
$so(10)$ spinor with ~$r_4=1$.
 We summarise the results in the following table:
\begin{equation}
\label{dim10a}
\begin{array}{lcrccr}
    \L_i && \dim(\L_i) && \chi=[r_1,r_2,r_3,r_4,r_5] & \dim(\chi)  \\
&&&&&\\
    \L_1 &&  16 && [0,0,0,0,1] &  16\\
\L_{15} &&  16 && [0,0,0,1,0] &  16\\
\L_2,\L_{14} &&  120 && [0,0,1,0,0] &  120\\
\L_3 &&  560 && [0,1,0,1,0] &  560\\
\L_{13} &&  560 && [0,1,0,0,1] &  560\\
\L_4,\L_{12}  &&  1820 && [1,0,1,0,0] &  945\\
\L_5  &&  4368 && [0,0,1,1,0] &  1200\\
\L_{11}  &&  4368 && [0,0,1,0,1] &  1200\\
\L_6,\L_{10} &&  8008 && [0,1,1,0,0] &  2970\\
\L_7  &&  11440  && [1,0,1,1,0] & 8800  \\
\L_9  &&  11440  && [1,0,1,0,1] & 8800  \\
\L_8 &&  12870  && [0,0,2,0,0] &  4125
\end{array}
\end{equation}
Part of the above analysis is done in the oscillator approach in
\cite{Gun}.

\bigskip

For odd $D$ the non-zero scalar products are:
~$(\g_j\,,\g_j) ~=~ 2\k$, ($j=1,\ldots,\ell-1$),
~$(\g_{\ell}\,,\g_{\ell}) ~=~ \k$,
~$(\g_j\,,\g_{j+1}) ~=~ -\k$, ($j=1,\ldots,\ell-1$).

In the case ~$D=9$, ~$\ell=4$, ~$n=16$, we have:
\eqnn{bas9}
&&\g_1=\a_5+2\a_6+\a_7+\a_9+2\a_{10}+\a_{11}\  , \cr
&&\g_2=\a_3+2\a_4+\a_5+\a_{11}+2\a_{12}+\a_{13}\  , \cr
&&\g_3=\a_2+\a_6+\a_7+2\a_8+\a_{9}+\a_{10}+\a_{14}\  , \cr
&&\g_4=\half(\a_1+\a_3+\a_{5}+\a_{7}+\a_{9}
+\a_{11}+\a_{13}+\a_{15})\  ,\eea
(with $\k=4$). Then, the $so(9)$ Dynkin labels
and dimensions are:
\eqnn{basm9}
&&r_1=a_5+2a_6+a_7+a_9+2a_{10}+a_{11}\  , \cr
&&r_2=a_3+2a_4+a_5+a_{11}+2a_{12}+a_{13}\  , \cr
&&r_3=a_2+a_6+a_7+2a_8+a_{9}+a_{10}+a_{14}\  , \cr
&&r_4=a_1+a_3+a_{5}+a_{7}+a_{9}
+a_{11}+a_{13}+a_{15}\  ,\eea
\eqn{dim9}
{\rm \dim}~L(r_1,...,r_4) ~ = ~
\frac{2^4\,n_1n_2n_3n_4}{7.5.3}\
\prod_{1\leq s<t \leq 4}
\frac{ n^2_t - n^2_s}{ (t -s) \ (t+s-1) } \ ,
\end{equation}
\eqnn{par9} &&n_1 ~=~ \half\,  (r_{4}+1) \ , \qquad
n_s ~=~ \half (r_4-1) + r_3 + \dots +
r_{5 - s} +s\ , ~~s=2,3,4, \qquad \\
&& n_{4} ~>~ n_3 ~>~ n_2 ~>~  n_1 > 0\ .\eea

The fundamental 16-dimensional $so(9)$ spinor - obtained
for ~$r_4=1$~ - is contained without reduction in both conjugated
16-dimensional fundamental ~$su(16)$~ UIRs with ~$a_1=1$~
and ~$a_{15}=1$.  We summarise the results in the following table:
\begin{equation}
\label{dim9a}
\begin{array}{lcrccr}
\L_i && \dim(\L_i) && \chi=[r_1,r_2,r_3,r_4] & \dim(\chi)  \\
&&&&&\\
\L_1,\L_{15} && 16 && [0,0,0,1] & 16 \\
\L_2,\L_{14} && 120 && [0,0,1,0] & 84  \\
\L_3,\L_{13}  && 560 && [0,1,0,1] & 432 \\
\L_4,\L_{12}  && 1 820 && [0,2,0,0] & 495 \\
\L_5,\L_{11}  && 4 368 && [1,1,0,1] & 2 560 \\
\L_6,\L_{10}  && 8 008 && [2,0,1,0] & 2 457 \\
\L_7,\L_{9}  && 11 440 && [1,0,1,1] & 5 040 \\
\L_8 && 12 870 && [0,0,2,0] & 1 980
\end{array}
\end{equation}

\bigskip

In the case ~$D=11$, ~$\ell=5$, ~$n=32$, we have:
\eqnn{bas11}
\g_1&=&\a_5+\a_{8}+\a_{10}+\a_{13}+\a_{14}+\a_{15}
+\cr &&+\a_{17}+\a_{18}+\a_{19}+\a_{22}+\a_{24}+\a_{27}
\ ,\cr
\g_2&=&\a_4 +\a_{7}+\a_{9}+\a_{10}+\a_{11}
+\a_{12}+\a_{14}+ 2\a_{15}+ 2\a_{16}+
\cr&&+2\a_{17}+\a_{18}+\a_{20}+\a_{21}
+\a_{22}+\a_{23}+\a_{25}+\a_{28}
\ ,\cr
\g_3&=&\a_3+\a_5+\a_{6}+ \a_8+\a_{9}+\a_{12}+
\a_{13}+\a_{14}+\a_{15}
+2\a_{16}+\cr&&+a_{17}+\a_{18}+\a_{19}+\a_{20}+
\a_{23}+\a_{24}+\a_{26}+\a_{27}+\a_{29}
\ ,\cr
\g_4&=&\a_2+\a_{4}+\a_6+\a_{7}+\a_{8} +\a_{10}
+\a_{11}+a_{12}+\a_{13}+\a_{14}+
\cr&&+  \a_{18}+\a_{19}+a_{20}+\a_{21}+
\a_{22}+\a_{24}+a_{25}+\a_{26}+\a_{28}+\a_{30}
\ ,\cr
\g_5&=&\half(\a_1+\a_3+\a_{5}+\a_{7}+\a_{9}
+\a_{11}+\a_{13}+\a_{15}+\cr&&+
\a_{17}+\a_{19}
+\a_{21}+\a_{23}+\a_{25}+\a_{27}+\a_{29}
+\a_{31}) \ ,\eea
(with $\k=8$). Then, the $so(11)$ Dynkin labels
and dimensions are:
\eqnn{basm11}
r_1&=&a_5+a_{8}+a_{10}+a_{13}+a_{14}+a_{15}
+\cr &&+a_{17}+a_{18}+a_{19}+a_{22}+a_{24}+a_{27}
\ ,\cr
r_2&=&a_4 +a_{7}+a_{9}+a_{10}+a_{11}
+a_{12}+a_{14}+ 2a_{15}+ 2a_{16}+
\cr&&+2a_{17}+a_{18}+a_{20}+a_{21}
+a_{22}+a_{23}+a_{25}+a_{28}
\ ,\cr
r_3&=&a_3+a_5+a_{6}+ a_8+a_{9}+a_{12}+
a_{13}+a_{14}+a_{15}
+2a_{16}+\cr&&+a_{17}+a_{18}+a_{19}+a_{20}+
a_{23}+a_{24}+a_{26}+a_{27}+a_{29}
\ ,\cr
r_4&=&a_2+a_{4}+a_6+a_{7}+a_{8} +a_{10}
+a_{11}+a_{12}+a_{13}+a_{14}+
\cr&&+  a_{18}+a_{19}+a_{20}+a_{21}+
a_{22}+a_{24}+a_{25}+a_{26}+a_{28}+a_{30}
\ ,\cr
r_5&=&a_1+a_3+a_{5}+a_{7}+a_{9}
+a_{11}+a_{13}+a_{15}+\cr&&+
a_{17}+a_{19}
+a_{21}+a_{23}+a_{25}+a_{27}+a_{29}
+a_{31} \ ,\eea
\eqn{dim11}
{\rm \dim}~L(r_1,...,r_5) ~ = ~ \frac{2^5\, n_1n_2n_3n_4n_5}
{9.7.5.3}\ \prod_{1\leq s<t \leq 5}
\frac{ n^2_t - n^2_s}{ (t -s) \ (t+s-1) } \ ,
\end{equation}
\eqnn{par11} &&n_1 ~=~ \half (r_5+1) \ ,
~~n_s ~=~ \half (r_5-1) + r_4 + \dots +
r_{6 -s} +s\ , ~~s=2,...,\ell, \qquad \\
&& n_5 ~>~ n_{4} ~>~ n_3 ~>~ n_2 ~>~  n_1 > 0\ . \eea

The fundamental 32-dimensional $so(11)$ spinor - obtained
for ~$r_5=1$~ - is contained without reduction in both
32-dimensional fundamental ~$su(32)$~ UIRs with ~$a_1=1$~
and ~$a_{31}=1$.  We summarise the results in the following table:
\begin{equation}
\label{dim11a}
\begin{array}{lcrccr}
    \L_i && \dim(\L_i) && \chi=[r_1,r_2,r_3,r_4,r_5] & \dim(\chi)  \\
&&&&&\\
    \L_1,\L_{31} &&  32 && [0,0,0,0,1] &  32\\
\L_2,\L_{30} &&  496 && [0,0,0,1,0] &  330\\
\L_3,\L_{29} &&  4 960 && [0,0,1,0,1] &  3 520\\
\L_4,\L_{28} &&  35 960 && [0,1,0,1,0] &  11 583 \\
\L_5,\L_{27} &&  201 376 && [1,0,1,0,1] &  28 512 \\
\L_6,\L_{26} &&  906 192 && [0,0,1,1,0] & 23 595 \\
\L_7,\L_{25} &&  3 365 856 && [0,1,0,1,1] &  160 160\\
\L_8,\L_{24} &&  10 518 300 && [1,0,1,1,0]  & 178 750 \\
\L_9,\L_{23} &&  28 048 800 && [0,1,1,0,1]  & 91 520 \\
\L_{10},\L_{22} &&  64 512 240 && [1,1,0,1,0] & 78 650 \\
\L_{11},\L_{21} &&  129 024 480 && [0,1,0,1,1] &  160 160 \\
\L_{12},\L_{20} &&  225 792 840 && [0,1,1,1,0] &  525 525 \\
\L_{13},\L_{19} &&  347 373 600 && [1,0,1,1,1] &  2 114 112 \\
\L_{14},\L_{18} &&  471 435 600 && [1,1,1,1,0] & 3 128 697  \\
\L_{15},\L_{17} &&  565 722 720 && [1,2,1,0,1] & 6 040 320  \\
\L_{16} &&  601 080 390 && [0,2,2,0,0] & 1 718 496
\end{array}
\end{equation}

\bigskip

{\bf Acknowledgement.} VKD and BSZ were supported in part by the
Bulgarian National Council for Scientific Research grant F-1205/02.

\end{document}